\documentclass[english, onecolumn, aps, prd, 10pt]{revtex4-2}
\usepackage[T1]{fontenc}
\usepackage[latin9]{inputenc}
\usepackage{units}
\usepackage{amsmath}
\usepackage{stackrel}
\usepackage{graphicx}

\makeatletter
\usepackage{braket}
\usepackage{blkarray}
\usepackage{graphicx}

\makeatother

\usepackage{babel}
\begin{document}
\title{If Mixed States Are Secretly Quickly Oscillating Pure States, Weak
Measurements Can Detect It}
\author{Igor Prlina}
\email{prlina@ipb.ac.rs}

\affiliation{Institute of Physics, University of Belgrade, Pregrevica 118, 11080 Belgrade, Serbia}
\begin{abstract}
The apparent nonunitary evolution in the black hole information paradox
and recent work on describing wavefunction collapse via nonunitary
nonlinear stochastic operators has motivated us to analyze whether
mixed states can be distinguished from quickly oscillating pure states.
We have demonstrated that the answer is no for all practical purposes
if only strong nonpostselected measurements are performed. However,
if weak measurements in postselected systems are used, mixed states
and quickly oscillating states produce different results. An experimental
procedure is proposed which could in principle determine the nature
of mixed states stemming from blackbody radiation, decoherence, thermalization
in solid state materials, Unruh radiation and Hawking radiation, among
others. The analysis in this work applies to all fast oscillations,
including those at Planck scale. As such, tabletop weak measurements
can be used to probe (very specific) potential high energy behavior,
where strong nonpostselected measurements cannot be applied. This
work also demonstrates that weak measurements are not equivalent to
a set of strong measurements without postselection since measurements
which are impossible for all practical purposes need to be excluded.
\end{abstract}
\maketitle

\section{Introduction}

One of the biggest unresolved problems in quantum theory is the so
called ``black hole information paradox'' \citep{InfParadox}. Due
to Hawking radiation \citep{Hawking}, one can have a pure state corresponding
to an isolated system collapsing into a black hole, which subsequently
fully evaporates, ending in a mixed thermal state. Such an evolution
seems to be non-unitary, which is against the postulates of quantum
dynamics. However, there is another important aspect of black hole
information paradox: an isolated system in equilibrium spontaneously
increases its entropy, which is forbidden by the laws of thermodynamics.
Without addressing the unitarity concerns, one possible resolution
to the thermodynamics issue would be if Hawking radiation is actually
a pure state, by definition of zero entropy, yet has properties which
make it practically indistinguishable from a mixed state. Of course,
pure states and mixed stated are vastly different, how can they be
practically indistinguishable? One possible answer is averaging over
time. This is relevant if the pure state is a superposition of vectors
whose relative phases oscillate very quickly in time.

In statistical physics, there is a long standing belief that averaging
a single system over long period of time gives the same probability
distribution as averaging over a large ensemble of systems in a given
moment of time. This is related to the ergodic theorem \citep{Ergodic}.
This equivalence is used in both statistical and quantum mechanics
to pick averaging over ensembles as the preferred method. However,
a consequence of the equivalence is that the two averagings must lead
to the same result. Namely, if one averages a quantum system over
ensemble and obtains a quantum state, one is allowed to average it
over time once more and the state must not change.

The averaging over time usually occurs because the experiment lasts
for some finite yet small amount of time. The study of continuous
measurements is quite complicated \citep{ContMeasure}. Instead of
continuous measurements, in this paper we will assume that all measurements
are perfect and instantaneous. That said, the exact moment that a
measurement occurred in cannot be precisely determined. Namely, any
device that observes time (say, a stopwatch) has a finite resolution.
This means that when we perform a set of measurements at some fixed
time, in actuality we are mixing measurement results from experiments
occurring in some interval $\Delta t$ around the specified moment
of time. This interval depends on the resolution of the time measurement
device. As such, the measured probability distributions do not correspond
to one single measurement at a specified time, but to a set of measurements
at different moments in time. Thus, the probabilities are effectively
averaged over time. We note that the same procedure applies even in
the case of continuous measurements, if these are modeled as a sequence
of instantaneous measurements. The averaging is not performed over
the duration of the continuous measurement, but over the time interval
in which the continuous measurement can start.

It has recently been suggested that the collapse of the wavefunction
in the act of measurement can be described using nonunitary, nonlinear
stochastic evolution \citep{ObjectiveCollapse}. Nonunitarity implies
that a pure state can evolve into an effectively mixed state, nonlinearity
implies that two identical vectors can evolve in different ways, and
stochastic evolution can be modeled by quick oscillations. Let us
elaborate on the claim that nonlinear evolution can lead to identical
vectors evolving in distinct ways. In quantum mechanics states differing
only by a complex multiplicative factor are usually considered to
be identical, $\ket{\psi}\equiv C\ket{\psi}.$ Since nonlinear evolution
in general does not have to be homogeneous, $E(t,C\ket{\psi})\neq CE(t,\ket{\psi}),$
the equivalence no longer holds. However, if we still treat all the
states that differ up to a complex factor as identical, the nonlinear
evolution will behave as if one state can evolve in different ways.
This analysis, as well as previous discussion, has lead us to ask
a question, the answer to which is the main goal of this paper: can
quickly oscillating states be distinguished from mixed states? As
we will demonstrate, without postselection, the answer is no for all
practical purposes. However, weak measurements in postselected systems
can distinguish the two possibilities. It is very important to note
that in this paper we do not claim that mixed states are fundamentally
pure quickly oscillating states, nor do we suggest a mechanism which
would cause such oscillations. We merely show that if such oscillations
exist, they can be detected using weak measurements.

As is well-known, quantum mechanics is indeterministic due to quantum
measurements, which can only be described in terms of probabilities.
In addition to absolute probabilities, in probability theory conditional
probabilities also exist. Conditional probabilities in quantum mechanics
correspond to postselected systems and are given by the Aharonov-Bergmann-Lebowitz
(ABL) rule \citep{ABL}. To measure an observable in a specified quantum
state, one must first prepare, that is, preselect a system. At some
initial time, a selective measurement is performed, and only the elements
of the ensemble that satisfy the preselection condition contribute
to the measurement result. Postselection is similar: after the measurement,
another selective measurement is performed, and the measurement results
corresponding to the elements of the ensemble which do not satisfy
the postselection results are discarded. The ABL probability does
not only depend on the eigenprojector corresponding to the eigenvalue
measured in the experiment. It also depends on all other eigenprojectors
corresponding to eigenvalues that could have been the result of the
measurement by the same apparatus, but haven't. This has lead to theoretical
discussions on the nature of objective probabilities and quantum counterfactuals
\citep{Threeboxes,CounterfactualBAD,CounterfactualGOOD}. A variant
of postselected systems has been used to describe the interaction
of ions with a conducting surface \citep{Rydberg,Nano}. The analysis
of postselected measurements under non-unitary evolution was performed
in \citep{NonunitaryABL}.

When one wants to determine what is the expectation value of a given
observable, one usually strongly couples it with the quantum system
and performs a strong measurement. This measurement provides more
information than the expectation value alone. The entire probability
distribution for all eigenvalues of the observable is obtained, and
the quantum state collapses in the act of measurement. One can calculate
the expectation value from the probability distribution, and ignore
the distribution itself. However, fundamentally the information was
available and as such the collapse has occurred. There is a way to
measure expectation values directly without collapsing the state.
Namely, one can choose to weakly couple the observable to the quantum
system. The measurement device is not capable of determining what
are individual measurement results of the given observable for each
member of the ensemble, however the mean value can be observed. This
type of measurement can be called weak non-postselected measurement.
When the experimental setup used to obtain weak non-postselected measurements
is used under a postselection condition, one obtains so-called weak
values of (postselected) weak measurements \citep{WeakMeasurement}.
Usually, using the term ``weak measurement'' implies that postselection
has been performed. Experimentally, weak measurements have been applied
in many different practical problems: to amplify the measurement signal
\citep{WeakValueAmplify}, to directly measure the wavefunction \citep{WavefunctionMeasure},
to measure ``trajectories'' in the double-slit experiment \citep{WeakTrajectories}
and many more. As we will demonstrate in this paper, weak measurements
can be used to distinguish quickly oscillating pure states from mixed
states as well.

This paper is organized as follows: In Section 2, we discuss averaging
over time and averaging over ensemble in quantum mechanics. In Section
3, we discuss weak measurements and weak values, and how they behave
under averaging over time. In Section 4, we discuss in greater detail
the case of two-state quantum systems. In Section 5, we discuss infinite
dimensional quantum systems, in both discrete and continuous basis.
In Section 6, an experimental protocol for distinguishing quickly
oscillating pure states from mixed states is described, as well as
the theoretical implications of the possibility of these measurements.
Finally, in Section 7, some concluding remarks about possible extensions
of the presented work are given.

\section{Averaging over Ensemble and over Time}

Let us begin by defining what we mean by ``quickly oscillating states''.
One can choose a pure state as some sum, such that each summand has
a time-dependent phase and a time independent real factor:
\begin{equation}
\ket{\psi}=A_{1}e^{i\varphi_{1}\left(t\right)}\ket{\psi_{1}}+A_{2}e^{i\varphi_{2}\left(t\right)}\ket{\psi_{2}}+...+A_{n}e^{i\varphi_{n}\left(t\right)}\ket{\psi_{n}}.\label{eq:QOStates}
\end{equation}
In the form of a statistical operator, this state can be written as
the projector

\begin{equation}
\varPi_{\psi}=\underset{i}{\sum}\left|A_{i}\right|^{2}\ket{\psi_{i}}\bra{\psi_{i}}+\underset{i\neq j}{\sum}A_{i}^{*}A_{j}e^{i\left(\varphi_{j}\left(t\right)-\varphi_{i}\left(t\right)\right)}\ket{\psi_{j}}\bra{\psi_{i}}.\label{eq:QOProj}
\end{equation}
When we say a state is quickly oscillating, we assume that the state
is (at least approximately for the duration of measurement) of the
form (\ref{eq:QOStates}), and that all the phase differences $\varphi_{j}\left(t\right)-\varphi_{i}\left(t\right),i\neq j,$
quickly change in time. If this projector is averaged over a time
interval $\Delta t$ corresponding to a finite resolution of the time
measurement device, such that this interval is long enough that the
phase differences change for a large number of multiples of $\pi$, one
obtains a mixed state:

\begin{equation}
\rho_{\psi}=\underset{i}{\sum}\left|A_{i}\right|^{2}\ket{\psi_{i}}\bra{\psi_{i}}.\label{eq:MixedState}
\end{equation}
A mixed state is obviously different than a pure state. However, states
are (usually) {[}9{]} not directly observed in quantum mechanics.
Instead, they are just a step towards obtaining measurable quantities,
that is, expectation values of observables, transition probabilities
and higher moments of probability distribution. The expectation value
of an observable $\hat{O}$ in state $\varPi_{\psi}$ is

\begin{equation}
\braket{\hat{O}}=\mathrm{Tr}\left[\varPi_{\psi}\hat{O}\right]=\underset{i}{\sum}\left|A_{i}\right|^{2}\braket{\psi_{i}|\hat{O}|\psi_{i}}+\underset{i\neq j}{\sum}A_{i}^{*}A_{j}e^{i\left(\varphi_{j}\left(t\right)-\varphi_{i}\left(t\right)\right)}\braket{\psi_{i}|\hat{O}|\psi_{j}}.\label{eq:QOExpectationValue}
\end{equation}
After averaging over the time interval $\Delta t$, one obtains the
same result as would have been obtained starting from the mixed state:

\begin{equation}
\braket{\hat{O}}_{t}\equiv\frac{1}{\Delta t}\stackrel[0]{\Delta t}{\int}\braket{\psi(t)|\hat{O}|\psi(t)}\mathrm{d}t=\braket{\hat{O}}\equiv\mathrm{Tr}\left[\rho_{\psi}\hat{O}\right]=\underset{i}{\sum}\left|A_{i}\right|^{2}\braket{\psi_{i}|\hat{O}|\psi_{i}}.\label{eq:TAExpectationValue}
\end{equation}
Similarly, if we introduce another state

\begin{equation}
\varPi_{a}=\ket{a}\bra{a},\label{eq:Proja}
\end{equation}
the transition probability from $\varPi_{\psi}$ to $\varPi_{a}$,

\begin{equation}
\mathrm{Tr}\left[\varPi_{\psi}\varPi_{a}\right]=\left|\braket{\psi|a}\right|^{2},\label{eq:TransProb}
\end{equation}
is equivalent to the expectation value of the projector $\varPi_{a}$
in the state $\ket{\psi}$. Since we have shown that under time averaging,
the state $\varPi_{\psi}$ gets replaced by the mixed state $\rho_{\psi}$
in the expression for any observable, including $\hat{O}=\varPi_{a}$,
we conclude that taking the time average of a transition probability
in a quickly oscillating state is equivalent to starting from the
corresponding mixed state:

\begin{equation}
\braket{\mathrm{Tr}\left[\varPi_{\psi}\varPi_{a}\right]}_{t}=\mathrm{Tr}\left[\rho_{\psi}\varPi_{a}\right]=\underset{i}{\sum}\left|A_{i}\right|^{2}\left|\braket{\psi_{i}|a}\right|^{2}.\label{eq:TATransProb}
\end{equation}

One can also consider time correlations:

\begin{equation}
C_{1}=\bra{\psi(t)}\hat{O}(\tau_{1})...\hat{O}(\tau_{N})\ket{\psi(t)}=\mathrm{Tr}\left[\varPi_{\psi}\hat{O}(\tau_{1})...\hat{O}(\tau_{N})\right].\label{eq:CorOfProd}
\end{equation}
The moments in time at which the observables are taken cannot be chosen
exactly due to the finite resolution of the time measurement device.
As such, in addition to averaging over time $t$, one must also average
over times $\tau_{i}$. If we do time averaging over $t$ first, due
to the fact that the product $\hat{O}(\tau_{1})...\hat{O}(\tau_{N})$
does not depend on time $t$, and due to the linearity of the trace,
one obtains

\begin{equation}
\braket{C_{1}}_{t}=\mathrm{Tr}\left[\rho_{\psi}\hat{O}(\tau_{1})...\hat{O}(\tau_{N})\right],\label{eq:TACorOfProd}
\end{equation}
that is, after time averaging over $t$, the quickly oscillating pure
state and the mixed state lead to the same correlator. This result
still needs to be averaged over times $\tau_{i}$ but since the intermediate
step is the same in both cases, the final result must be identical
as well. One can also evaluate the product of means at different moments
of time:

\begin{equation}
C_{2}=\bra{\psi(t+\tau_{1})}\hat{O}\ket{\psi(t+\tau_{1})}...\bra{\psi(t+\tau_{N})}\hat{O}\ket{\psi(t+\tau_{N})},\label{eq:ProdOfCor}
\end{equation}
where we need to average over both $t$ and $\tau_{i}$ as before.
Each mean depends only on a single time $\tau_{i}$, and when averaging
over it, time $t$ is treated as a constant phase. As such, using
the same arguments as in (\ref{eq:TAExpectationValue}),

\begin{equation}
\bra{\psi(t+\tau_{i})}\hat{O}\ket{\psi(t+\tau_{i})}_{\tau_{i}}=\mathrm{Tr}\left[\rho_{\psi}\hat{O}\right],\label{eq:TAProdOfCor}
\end{equation}
which does not depend on time. Thus, the product of means after averaging
is the same for a quickly oscillating pure state and for a corresponding
mixed state. Note that this analysis also applies for the special
case where all parameters $\tau_{i}$ are equal to zero, which corresponds
to the higher moments of probability distribution. Due to the finite
resolution of the time measurement device, one cannot be certain that
they are correlating simultaneous experimental results, and not with
results in slightly different moments of time. 

As such, one can substitute all projectors onto states of quickly
oscillating phases with corresponding mixed states and omit averaging
over time. No realistic measurement result will be modified in this
way. The quickly oscillating state and the mixed state are fully equivalent
for observable quantities, and as such, the question whether the state
is fundamentally pure and quickly oscillating, or mixed without rapid
time dependence, appears to be metaphysical for all practical purposes.
In the following section we will show that this is not the case if
one considers weak measurements on postselected systems. Finally,
let us note that the previous analysis also applies to the case of
continuous measurements as long as they are modeled as a sequence
of instantaneous measurements. The first measurement in the sequence
collapses the quickly oscillating state into a state that changes
slowly in time. Thus, after the initial moment of measurement, quick
oscillations have no effect on the experimental result. That said,
the initial moment of measurement cannot be precisely controlled just
like in the single instantaneous measurement case, and the experimental
result needs to be averaged over the resolution of the time measurement
device (not over the duration of the continuous measurement).

\section{Weak Values under Averaging over Time}

In this Section, we will focus on weak measurements. Unlike in non-postselected
case, weak measurements in postselected systems do not simply give
an expectation value. The postselection introduces a second state
into consideration, which can be evolved backwards in time towards
the moment of the weak measurement. As such, the weak measurement
will depend on the matrix element of the observable indexed by the
two states. The weak value of an observable $\mathrm{\hat{O}}$ in
a quantum system described by the preselected and postselected states
$\ket{\varPsi_{1}}$ and $\ket{\varPsi_{2}}$ respectively, is, as
shown in {[}3{]}, given by 

\begin{equation}
O_{w}=\frac{\braket{\varPsi_{1}|\mathrm{\hat{O}}|\varPsi_{2}}}{\braket{\varPsi_{1}|\varPsi_{2}}}.\label{eq:WVPure}
\end{equation}
The pure states $\ket{\varPsi_{1}}$ and $\ket{\varPsi_{2}}$ correspond
to statistical operators $\hat{\varPi}_{1}=\ket{\varPsi_{1}}\bra{\varPsi_{1}}$
and $\hat{\varPi}_{2}=\ket{\varPsi_{2}}\bra{\varPsi_{2}}$ respectively.
We can rewrite the expression using the statistical operators

\begin{equation}
O_{w}=\frac{\mathrm{Tr}\left[\hat{\varPi}_{1}\hat{\mathrm{O}}\hat{\varPi}_{2}\right]}{\mathrm{Tr}\left[\hat{\mathit{\varPi}}_{1}\hat{\varPi}_{2}\right]},\label{eq:WVProj}
\end{equation}
with a simple proof

\[
\frac{\mathrm{Tr}\left[\hat{\varPi}_{1}\hat{\mathrm{O}}\hat{\varPi}_{2}\right]}{\mathrm{Tr}\left[\hat{\mathit{\varPi}}_{1}\hat{\varPi}_{2}\right]}=\frac{\mathrm{Tr}\left[\ket{\varPsi_{1}}\bra{\varPsi_{1}}\hat{\mathrm{O}}\ket{\varPsi_{2}}\bra{\varPsi_{2}}\right]}{\mathrm{Tr}\left[\ket{\varPsi_{1}}\braket{\varPsi_{1}|\varPsi_{2}}\bra{\varPsi_{2}}\right]}=\frac{\bra{\varPsi_{1}}\hat{\mathrm{O}}\ket{\varPsi_{2}}\braket{\varPsi_{2}|\varPsi_{1}}}{\braket{\varPsi_{1}|\varPsi_{2}}\braket{\varPsi_{2}|\varPsi_{1}}}.
\]
As can be seen from equation (\ref{eq:WVPure}), weak values are complex
numbers. The real part of a weak value is directly observable as the
position of the pointer of the measurement device, and the imaginary
part corresponds to the momentum of the pointer \citep{WeakMeasurement}.
Since the pointer can be a macroscopic object, the uncertainty principle
can be ignored. An experimental procedure which directly observes
the modulus and the phase of the weak value based on a quantum eraser
has been developed as well \citep{Eraser}.

If the phases of the states quickly oscillate in time, the observable
weak value needs to be time-averaged. As such, the measurement result
directly corresponds to

\begin{equation}
\braket{O_{w}}_{t}=\langle\frac{\mathrm{Tr}\left[\hat{\varPi}_{1}\hat{\mathrm{O}}\hat{\varPi}_{2}\right]}{\mathrm{Tr}\left[\hat{\mathit{\varPi}}_{1}\hat{\varPi}_{2}\right]}\rangle_{t}\label{eq:TAWVProj}
\end{equation}
where 
\begin{equation}
\braket{O_{w}}_{t}\equiv\frac{1}{T}\stackrel[0]{T}{\int}O_{w}\mathrm{d}t,\label{eq:TAdef}
\end{equation}
and $T$ is the period of fast oscillations. In the previous section
we presented standard arguments that a quickly oscillating state $\varPi_{1}$
can be substituted by the corresponding mixed state $\rho_{1}$. If
we do that, we obtain

\begin{equation}
O_{w}=\frac{\mathrm{Tr}\left[\hat{\rho}_{1}\hat{\mathrm{O}}\hat{\varPi}_{2}\right]}{\mathrm{Tr}\left[\hat{\mathit{\rho}}_{1}\hat{\varPi}_{2}\right]}=\frac{\mathrm{\braket{\mathrm{Tr}\left[\hat{\varPi}_{1}\hat{\mathrm{O}}\hat{\varPi}_{2}\right]}_{t}}}{\braket{\mathrm{Tr}\left[\hat{\mathit{\varPi}}_{1}\hat{\varPi}_{2}\right]}_{t}}.\label{eq:WVoneMixed}
\end{equation}
Here we make an important observation: if the state is fundamentally
quickly oscillating and pure, the value of the weak measurement will
differ from the case when the state is fundamentally mixed. The question
regarding the nature of the state is no longer metaphysical, it becomes
experimentally testable.

The expression (\ref{eq:WVoneMixed}) is not just a result of the
aforementioned substitution. It applies for systems which are preselected
in mixed states and can be derived from (\ref{eq:WVProj}). To do
so, let us explain what it means to preselect or postselect a system
into a mixed state. Mixed states are described by statistical operators:
hermitian operators with non-negative eigenvalues, with unit trace.
That means that the preselected and postselected states can be written
as

\begin{equation}
\rho_{i}=\underset{i}{\sum}p_{i}\hat{\Pi}_{i},\qquad\rho_{f}=\underset{j}{\sum}q_{j}\hat{\Pi}_{j}.\label{eq:MixedDecomp}
\end{equation}
Preselecting into the mixed state $\rho_{i}$ can be done by utilizing
different preselection criteria for different members of the ensemble:
$p_{i}$ is the ratio of members of the ensemble that are preselected
in the pure state $\hat{\Pi}_{i}$. Similar interpretation applies
for postselection. Thus, a weak measurement with mixed state can be
considered as a combination of weak measurements with pure states.
The probability of a random member of the ensemble corresponding to
the pure states $\hat{\Pi}_{i}$ and $\hat{\Pi}_{j}$ is $p_{i}q_{j}$.
However, not all members of the ensemble satisfy the postselection
criterion. Thus, the former probability needs to be multiplied by
the probability that the postselection is satisfied, which is the
transition probability from the initial pure state to the final pure
state, $\mathrm{Tr}\left[\hat{\mathit{\varPi}}_{i}\hat{\varPi}_{j}\right]$.
As such, each weak measurement corresponding to a pure state pair
$\hat{\mathit{\varPi}}_{i}$, $\hat{\mathit{\varPi}}_{j}$ is weighted
by a factor of $p_{i}q_{j}\mathrm{Tr}\left[\hat{\mathit{\varPi}}_{i}\hat{\varPi}_{j}\right]$,
normalized by the total probability of postselection occurring, $\underset{k,l}{\sum}p_{k}q_{l}\mathrm{Tr}\left[\hat{\mathit{\varPi}}_{k}\hat{\varPi}_{l}\right]$.
Consequently, the weak value is

\begin{equation}
O_{w}=\underset{i,j}{\sum}\frac{p_{i}q_{j}\mathrm{Tr}\left[\hat{\mathit{\varPi}}_{i}\hat{\varPi}_{j}\right]}{\underset{k,l}{\sum}p_{k}q_{l}\mathrm{Tr}\left[\hat{\mathit{\varPi}}_{k}\hat{\varPi}_{l}\right]}\frac{\mathrm{Tr}\left[\hat{\varPi}_{i}\hat{\mathrm{O}}\hat{\varPi}_{j}\right]}{\mathrm{Tr}\left[\hat{\mathit{\varPi}}_{i}\hat{\varPi}_{j}\right]}.\label{eq:WVMixedBigSum}
\end{equation}
The trace in the numerator of the first fraction cancels with the
trace in the denominator of the second fraction. Using the linearity
of the trace, and equation (\ref{eq:MixedDecomp}), we obtain:

\begin{equation}
O_{w}=\frac{\mathrm{Tr}\left[\hat{\rho}_{1}\hat{\mathrm{O}}\hat{\rho}_{2}\right]}{\mathrm{Tr}\left[\hat{\mathit{\rho}}_{1}\hat{\rho}_{2}\right]}.\label{eq:WVmixed}
\end{equation}

It remains to answer how is it possible that direct averaging over
time and substitution of the pure state into the mixed state no longer
give same results. Does this imply that in the case of weak measurements,
the equivalence of averaging a system over time and averaging a system
over the ensemble of all possible configurations no longer holds?
Luckily, the answer is no. As described, in the previous analysis
weak measurements are performed in some time interval corresponding
to the finite resolution of the time measurement device. Within this
interval, the weak measurements are uniformly distributed. A mixed
state corresponds to averaging the pure state over the ensemble of
all possible configurations. Each configuration is equally present
in the mixture. The different configurations in the ensemble are described
by different phases, which correspond to different moments in time.
However, not all members of the initial ensemble will satisfy the
postselection condition at the same rate: different phases have different
transition probabilities. By using the mixed state, we overcount the
phases which are more likely to survive postselection. This would
correspond to a set of weak measurements which are not uniformly distributed
within the time resolution of the time measurement device. To compensate,
the expression needs to be weighted by the probability of satisfying
the postselection condition, leading to the correct result. The averaging
needs to be performed over the postselected ensemble, not the initial
one, and as such, the mixed state corresponding to the averaging over
the initial ensemble cannot be directly used in the case of quickly
oscillating states. This problem is not present when the state is
fundamentally mixed, because the postselection probability is the
same for all members of the subensembles corresponding to given pure
states in the mixture.

\section{Two-State Quantum Systems}

In this Section we will focus on the case of two-state vector systems.
We will label these two states $\ket{+}$ and $\ket{-}$. We introduce
a quickly oscillating state $\ket{\psi_{1}}$ corresponding to the
preselection condition, and a state $\ket{\psi_{2}}$ corresponding
to postselection which changes slowly with time in the measurement
interval:

\begin{equation}
\ket{\psi_{1}}=N_{1}(\ket{+}+Ae^{i\varphi}\ket{-}),\qquad\ket{\psi_{2}}=N_{2}(\ket{+}+Be^{i\chi}\ket{-}).\label{eq:QO2State}
\end{equation}
The coefficients $A$ and $B$ are taken to be strictly positive.
The negative signs can be absorbed into the phases. The coefficients
$N_{i}$ are normalization factors, $N_{1}=\left(1+A^{2}\right)^{-\nicefrac{1}{2}}$,
$N_{2}=\left(1+B^{2}\right)^{-\nicefrac{1}{2}},$ but since the expression
for weak value is linear in both the preselected and the postselected
state, both in the numerator and the denominator, the normalization
factors will cancel out in the expression for weak values. This is
relevant for the next Section. The weak value of observable $\hat{O}$
for quickly oscillating state $\ket{\psi_{1}}$ becomes

\begin{equation}
O_{w}=\frac{\braket{+|\hat{O}|+}+Be^{i\chi}\braket{+|\hat{O}|-}+Ae^{-i\varphi}\braket{-|\hat{O}|+}+ABe^{i\left(\chi-\varphi\right)}\braket{-|\hat{O}|-}}{1+ABe^{i\left(\chi-\varphi\right)}}.\label{eq:WVQO2State}
\end{equation}
With a proper choice of the measured observable, the expression can
simplify significantly. For example, let us pick the polarization
observable:

\begin{equation}
\hat{O}=\ket{+}\bra{+}-\ket{-}\bra{-}\equiv\hat{S.}\label{eq:SpinDef}
\end{equation}
Its weak value is given by the expression

\begin{equation}
S_{w}=\frac{1-ABe^{i\left(\chi-\varphi\right)}}{1+ABe^{i\left(\chi-\varphi\right)}}.\label{eq:SpinWVQO2State}
\end{equation}
Weak values are complex numbers, with both the real and the imaginary
part being directly measurable. After some manipulation, it can be
shown that the real part of this weak value is

\begin{equation}
\mathrm{Re}\left[S_{w}\right]=\frac{\frac{1-A^{2}B^{2}}{2AB}}{\frac{1+A^{2}B^{2}}{2AB}+\cos{\left(\chi-\varphi\right)}}.\label{eq:ReSpinWVQO2State}
\end{equation}
In order to explicitly evaluate the time average, we will assume that
the phase difference depends linearly on time, with a very high frequency:

\begin{equation}
\chi-\varphi=\omega t+\phi,\qquad\phi=const.\label{eq:PhaseDiff2State}
\end{equation}
Since the weak value is a periodic function, the average over a time
interval much larger than the period of oscillations is equal to the
average over a single period:

\begin{equation}
\braket{\mathrm{Re}\left[S_{w}\right]}_{t}=\frac{1}{T}\stackrel[0]{T}{\int}\mathrm{Re}\left[S_{w}\right]\mathrm{d}t.\label{eq:TAdefRe}
\end{equation}
Using the following known integral

\begin{equation}
\stackrel[0]{2\pi}{\int}\frac{dx}{a+\cos{x}}=\frac{2\pi}{\sqrt{a^{2}-1}},\qquad a>1,\label{eq:IntDef}
\end{equation}
 we can evaluate the average: 
\begin{equation}
\braket{\mathrm{Re}\left[S_{w}\right]}_{t}=\mathrm{sgn}\left[1-A^{2}B^{2}\right],\label{eq:TAReSpin2State}
\end{equation}
that is, the observed real part of the weak value can only ever be
either +1 or -1, other than the special case of $AB=1$, when the
expression (\ref{eq:IntDef}) does not hold, but it can be seen from
(\ref{eq:ReSpinWVQO2State}) that the real part of the weak value
is exactly zero in that case. It is important to note that this result
does not depend on the frequency of oscillations, as long as the oscillation
period is much shorter that the duration of the measurement. The imaginary
part of the weak value is given by the expression:

\begin{equation}
\mathrm{Im}\left[S_{w}\right]=-\frac{2AB\sin{\left(\chi-\varphi\right)}}{1+A^{2}B^{2}+2AB\cos{\left(\chi-\varphi\right)}}.\label{eq:ImSpin2State}
\end{equation}
 If we again assume linear dependence of the phase difference on time,
the time averaged imaginary part of the weak value becomes zero, since
we are averaging an odd function over its period:

\begin{equation}
\braket{\mathrm{Im}\left[S_{w}\right]}_{t}=0.\label{eq:TAImSpin2State}
\end{equation}
Now we will consider the weak value of the same observable when the
quickly oscillating pure state is substituted by the corresponding
mixed state:

\begin{equation}
\rho_{1}=N_{1}^{2}(\ket{+}\bra{+}+A^{2}\ket{-}\bra{-}),
\end{equation}
\[
\Pi_{2}=\ket{\psi_{2}}\bra{\psi_{2}}=N_{2}^{2}(\ket{+}\bra{+}+B^{2}\ket{-}\bra{-}+Be^{i\chi}\ket{-}\bra{+}+Be^{-i\chi}\ket{+}\bra{-}).
\]
The weak value is now

\begin{equation}
S_{w}=\frac{1-A^{2}B^{2}}{1+A^{2}B^{2}}.\label{eq:SpinWVmixed2State}
\end{equation}
Note that the same result holds even when we postselect the system
in the corresponding mixed state $\rho_{2}=N_{2}^{2}(\ket{+}\bra{+}+B^{2}\ket{-}\bra{-}).$
We see that this expression corresponds to averaging the numerator
and the denominator of expression (\ref{eq:ReSpinWVQO2State}) separately.
However, such an averaging is not proper. Just like in the quickly
oscillating pure states case, the imaginary part is zero. However,
the real part is significantly different. The weak value can take
on any value in the range $[-1,1]$, depending on the choice of A
and B, and is not limited to +1, -1 and 0 like in the quickly oscillating
pure state case. As such, we can see that weak measurement can be
used to distinguish whether states are quickly oscillating but fundamentally
pure, or if the states are fundamentally mixed. 

\section{Quantum Systems of Arbitrary Number of Dimensions}

The previous analysis can easily be repeated in the case of countably
many dimensions. Now, the quickly oscillating pure state is

\begin{equation}
\ket{\psi_{1}}=\underset{j}{\sum}\widetilde{A}_{j}e^{i\varphi_{j}}\ket{j}\label{eq:QOInfState}
\end{equation}
and the corresponding mixed state is
\begin{equation}
\rho_{1}=\underset{j}{\sum}\left|\widetilde{A}_{j}\right|^{2}\ket{j}\bra{j}.\label{eq:MixedInfState}
\end{equation}
We choose two basis vectors, $\ket{a}$ and $\ket{b}$, and we measure
the observable
\begin{equation}
\hat{O}=\ket{a}\bra{a}-\ket{b}\bra{b},\label{eq:InfSpin}
\end{equation}
while for the postselected state we choose
\begin{equation}
\ket{\psi_{2}}=N_{2}(\ket{a}+Be^{i\chi}\ket{b}).\label{eq:PostStateInf}
\end{equation}
We can rewrite the states as
\begin{equation}
\ket{\psi_{1}}=N_{1}(\ket{a}+Ae^{i\varphi}\ket{b}+\underset{j\neq a,b}{\sum}A_{j}e^{i\varphi_{j}}\ket{j})\label{eq:QOInfStateNorm}
\end{equation}
\[
\rho_{1}=\left|N_{1}\right|^{2}(\ket{a}\bra{a}+A^{2}\ket{b}\bra{b}+\underset{j\neq a,b}{\sum}A_{j}^{2}\ket{j}\bra{j}).
\]
The coefficients $N_{1}$ and $N_{2}$ are normalization factors which
will cancel out in the expression for the weak value. The chosen observable
$\hat{O}$ is non-zero only in a two-dimensional subspace. Both the
initial and final state have the same form as in the two-dimensional
case when projected to the said subspace. As such, it is easy to see
that the results from the two-dimensional case are obtained once more
for countably many dimensions, both in finite and infinite cases.

In the case of a continuous basis, the analysis requires more finesse.
A quickly oscillating state is taken to be of the following form:
\begin{equation}
\ket{\psi_{1}}=\stackrel[-\infty]{+\infty}{\int}A(x)e^{i\varphi(x,t)}\ket{x}\mathrm{d}x,\label{eq:QOStateCont}
\end{equation}
with the corresponding projector
\begin{equation}
\Pi_{1}=\ket{\psi_{1}}=\stackrel[-\infty]{+\infty}{\int}\stackrel[-\infty]{+\infty}{\int}A(x)A(y)e^{i\left[\varphi(x,t)-\varphi(y,t)\right]}\ket{x}\bra{y}\mathrm{d}x\mathrm{d}y.\label{eq:QOProjCont}
\end{equation}
We will assume that $A(x)$ and $\varphi(x,t)$ are smooth functions.
However, like in the previous case, we will take $A(x)$ to be non-negative,
and for originally negative values, we will absorb $\pi$ into the
phase. This will make $\varphi(x,t)$ piecewise continuous and piecewise
differentiable. In order for $\Pi_{1}$ to average over time into
the desired effectively mixed state, the off-diagonal elements ought
to average to zero. This will be the case as long as $e^{i\left[\varphi(x,t)-\varphi(y,t)\right]}$
is quickly oscillating in time. However, if we take $x$ and $y$
to be arbitrarily close, the oscillations will stop being fast at
some point. As such, time averaging of this quickly oscillating state
will never give an exact mixed state. That said, just like for time,
there is a finite resolution of realistic measurement. There exists
a $\Delta x$ such that for all practical purposes, $x$ and $x+\Delta x$
are experimentally indistinguishable. As such, matrix elements $\ket{x}\bra{x}$
and $\ket{x}\bra{x+\Delta x}$ are effectively identical. Thus, the
projector $\Pi_{1}$ averages out over time into a matrix for all
practical purposes indistinguishable from the mixed state

\begin{equation}
\rho_{1}=\stackrel[-\infty]{+\infty}{\int}A^{2}(x)\ket{x}\bra{x}\mathrm{d}x,\label{eq:MixedStateProj}
\end{equation}
as long as $e^{i\left[\varphi(x,t)-\varphi(y,t)\right]}$ is a quickly
oscillating function in time for any $x$, and any $y>x+\Delta x$.
For simplicity, we can assume

\begin{equation}
\varphi(x,t)=-\omega(x)t+\varphi(x).\label{eq:PhaseDiffCont}
\end{equation}
For the averaging to hold, the following relationship must be true
for all $x$:
\begin{equation}
\left|\omega'(x)\right|\gg\frac{2\pi}{\Delta x\Delta t},\label{eq:FreqRateCond}
\end{equation}
where $\omega'(x)$ is the first derivative of $\omega(x)$, and $\Delta t$
and $\Delta x$ are the state of the art experimental resolutions
of time and the observable with a continuous spectrum chosen as the
basis, respectively. Due to the finite resolution of the observable
$\hat{X}$, the effective mixed state might occur even at a fixed
moment of time, due to averaging over $x$. This will happen if $\varphi(x)$
is a quickly oscillating function in $x$. Otherwise, we will assume
that $\varphi(x)$ is slowly changing with $x$.

As before, we can choose an observable and postselected state such
that the quickly oscillating state and the corresponding mixed state
can be distinguished via a weak measurement. For the observable, we
will pick
\begin{equation}
\hat{O}=\stackrel[a-\Delta a]{a}{\int}\ket{x}\bra{x}\mathrm{d}x-\stackrel[a]{a+\Delta a}{\int}\ket{x}\bra{x}\mathrm{d}x,\label{eq:SpinCont}
\end{equation}
and for the postselected state we will take
\begin{equation}
\ket{\psi_{2}}=\stackrel[a-\Delta a]{a}{\int}B(x)e^{i\chi(x)}\ket{x}\mathrm{d}x+\stackrel[a]{a+\Delta a}{\int}B(x)e^{i\chi(x)}\ket{x}\mathrm{d}x.\label{eq:PostStateCont}
\end{equation}
For the weak value we obtain
\begin{equation}
O_{w}=\frac{\stackrel[a-\Delta a]{a}{\int}A(x)B(x)e^{i\left[\chi(x)-\varphi(x,t)\right]}\mathrm{d}x-\stackrel[a]{a+\Delta a}{\int}A(x)B(x)e^{i\left[\chi(x)-\varphi(x,t)\right]}\mathrm{d}x}{\stackrel[a-\Delta a]{a}{\int}A(x)B(x)e^{i\left[\chi(x)-\varphi(x,t)\right]}\mathrm{d}x+\stackrel[a]{a+\Delta a}{\int}A(x)B(x)e^{i\left[\chi(x)-\varphi(x,t)\right]}\mathrm{d}x}.\label{eq:WVQOContState}
\end{equation}
In order to evaluate this expression, we need to make assumptions
about the phase of the quickly oscillating state, as well as choose
suitable values of parameters of the postselected state. In what follows
we will assume that the quickly changing phase is
\begin{equation}
\varphi(x,t)=-\Omega xt+\Phi x,\label{eq:PhaseContFull}
\end{equation}
where $\Omega$ and $\Phi$ are constants. This choice of phase ensures
that the averaging over time is valid as long as $\left|\Omega\right|=\left|\omega'(x)\right|$
satisfies the expression (\ref{eq:FreqRateCond}). If the parameter
$\Phi$ is large, averaging over $x$ is also allowed, and in the
case that $\varphi\left(x\right)$ is a slowly changing function,
we approximate it with the linear term. For simplicity, we will choose
the phase of the postselected state to always be zero. The amplitude
of the quickly oscillating state, $A(x)$, cannot be controlled, but
it can be measured without postselection since $A^{2}(x)$ is the
probability of finding the effectively mixed state in state $\ket{x}.$
As such, we will treat $A(x)$ as a known function, and use it in
our choice of the postselected state. For the function $B(x)$ of
the postselected state, we will choose
\begin{equation}
B(x)=\frac{NC_{1}}{A(x)},\quad a-\Delta a<x<a,\quad B(x)=\frac{NC_{2}}{A(x)},\quad a<x<a+\Delta a,\label{eq:PostStateContFull}
\end{equation}
where $N$ is the normalization factor, and $C_{1}$ and $C_{2}$
are positive constant parameters. Now we can evaluate the expression
(\ref{eq:WVQOContState}):

\begin{equation}
O_{w}=\frac{C_{1}+C_{2}-C_{1}e^{-i(\Omega\Delta at-\Phi\Delta a)}-C_{2}e^{i(\Omega\Delta at-\Phi\Delta a)}}{C_{1}-C_{2}-C_{1}e^{-i(\Omega\Delta at-\Phi\Delta a)}+C_{2}e^{i(\Omega\Delta at-\Phi\Delta a)}},\label{eq:WVContIntx}
\end{equation}
and it is simple to show that this expression is equivalent to
\begin{equation}
O_{w}=\frac{1-\frac{C_{2}}{C_{1}}e^{i(\Omega\Delta at-\Phi\Delta a)}}{1+\frac{C_{2}}{C_{1}}e^{i(\Omega\Delta at-\Phi\Delta a)}}.\label{eq:WVContIntxSimple}
\end{equation}
The last expression has the same form as the expression (\ref{eq:SpinWVQO2State}),
and as such, gives analogous result after time averaging:

\begin{equation}
\langle O_{w}\rangle_{t}=\mathrm{sgn}[1-\frac{C_{2}^{2}}{C_{1}^{2}}].\label{eq:TAWVQOCont}
\end{equation}
In the case that the initial state is mixed, the weak value becomes

\begin{equation}
O_{w}=\frac{\stackrel[a-\Delta a]{a}{\int}A^{2}(x)B^{2}(x)dx-\stackrel[a]{a+\Delta a}{\int}A^{2}(x)B^{2}(x)dx}{\stackrel[a-\Delta a]{a}{\int}A^{2}(x)B^{2}(x)dx+\stackrel[a]{a+\Delta a}{\int}A^{2}(x)B^{2}(x)dx}\label{eq:WVMixedCont}
\end{equation}
and after applying the condition (\ref{eq:PostStateContFull}), the
expression evaluates into
\begin{equation}
O_{w}=\frac{1-\frac{C_{2}^{2}}{C_{1}^{2}}}{1+\frac{C_{2}^{2}}{C_{1}^{2}}}.\label{eq:WVMixedContFull}
\end{equation}
As such, weak measurements can be used to test the nature of effectively
mixed states even in the continuous case.

\section{Possible Experiments}

In Section 3 we have explained how to preselect a system in a mixed
quantum state: use different preselection criteria on different members
of the ensemble. However, this is not the approach we suggest in potential
experiments. We have demonstrated that weak measurements can be used
to distinguish mixed states from quickly oscillating ones. As such,
we should not be preselecting the states ourselves. Instead, we introduce
a source of states which should be mixed according to theory. These
can be experimentally feasible, like a radiating black body, a quantum
state prepared long time ago likely to have experienced decoherence,
or electrons in some material. We can also consider thought experiments
involving Unruh radiation \citep{Unruh,UnruhReview} or Hawking radiation.
In Figure 1, we give a sketch of such experiments. For sources of
continuous states, two experiments need to be conducted: a strong
measurement in nonpostselected system which gives the amplitude of
the effective mixed state used to choose the proper postselected state,
as well as a weak measurement on a postselected system. We are not
aware of any strong arguments why the easy-to-measure states involved
in thermal radiation, decoherence, or condensed matter should be quickly
oscillating. However, we still suggest conducting the corresponding
experiments, since they should not require significant investment
for groups which are already experimentally utilizing weak measurements.
This is especially true in the finite dimensional case, given that
weak measurements of polarization are relatively common. Weak measurements
of Unruh and Hawking radiation are practically unfeasible, but possible
in principle. These thought experiments are primarily relevant for
theoretical considerations.

\includegraphics{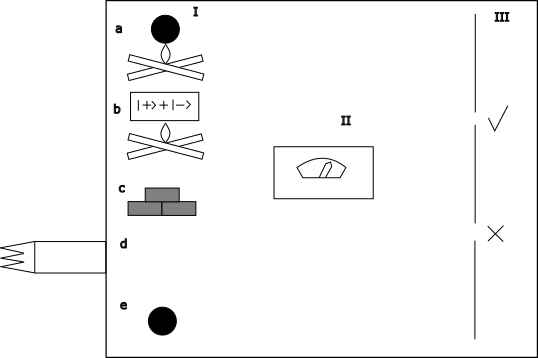}
\begin{center}
Figure 1. An abstracted sketch of possible experiments. Column I represents
possible sources of mixed states. Object II is the weak measurement
device. Object III is the postselection measurement. The represented
sources of mixed states are as follows. a) a black body producing
thermal radiation b) a pure state producing a mixed state via decoherence
c) a solid state material described by a mixed state d) a source of
constant acceleration leading to Unruh radiation, represented by a
rocket engine accelerating the entire experimental setup e) a black
hole producing Hawking radiation
\par\end{center}

We do not claim that the quick oscillations exist, nor do we suggest
that the quick oscillations occur at Planckian frequencies. However,
it is possible that they are indeed present in nature and that they
occur at Planck scale. This on it's own is of some importance. Namely,
if mixed states are fundamentally pure states oscillating at Planckian
frequencies, tabletop weak measurements would be able to observe the
effect. As such, we have shown that there exist possible Planck scale
phenomena which are observable by weak measurements in postselected
systems, while invisible under strong nonpostselected measurements.
As such, we suggest there is merit in further, more rigorous study
of weak measurements in the framework of quantum field theory, since
other applications of weak measurements could be found.

It has been argued that weak measurements are equivalent to a set
of nonpostselected strong measurements \citep{Weak=00003DStrong}.
As such, weak measurements would contain no new information relative
to nonpostselected measurements. As we have shown in this work, there
are experimental questions that cannot be answered by strong postselected
measurements, but can be investigated via weak measurements. This
occurs because there are measurements which are impossible for all
practical purposes. In our case, measurements didn't have perfect
time resolution. It is obvious that a perfect measurement occurring
at a fully fixed moment of time can distinguish a quickly oscillating
state from a mixed state. However, as explained, such a measurement
cannot be performed in a realistic setting. A feasible weak measurement
might be equivalent to a set of strong measurements on nonpostselected
systems, but such that some of those equivalent measurements are not
possible for all practical purposes. Thus, in practice, weak measurements
may lead to new information. 

\section{Concluding Remarks}

The question if states are fundamentally mixed, or they are pure but
with relative phases quickly oscillating is not metaphysical. This
question can be answered using weak measurements in postselected systems.
Mixed states can be obtained in multiple ways, for example: by merging
multiple ensembles together, by performing a strong nonselective quantum
measurement, by thermalization, by decoherence, by accelerating the
system, by observing a radiating black hole. We suggest that applying
weak measurements on these states can determine their true nature.
Some of these experiments should not be difficult nor expensive to
do, while some are merely thought experiments. While measuring Hawking
radiation from black holes is definitely beyond current experimental
reach, the presented results show that, in principle, weak measurements
can be used to see if black hole radiation is pure or mixed. This
might be of relevance for the so-called black hole information paradox,
which questions how a pure state can evolve into a mixed state after
evaporating from a black hole. This work has also demonstrated that
weak measurements are not equivalent to a set of strong measurements
without postselection when measurements which are impossible for all
practical purposes are excluded.

There are multiple possible extensions of presented work. It is possible
to use weak measurements to find decoherence rate in different systems.
It might also be possible to generalize the framework of weak measurements
to quantum field theory, and try to analyze the quantum information
paradox with more rigor, as well as look for other questions weak
measurements can answer that nonpostselected measurements cannot.
Additionally, weak measurements might be used to test different models
of objective wavefunction collapse which depend on stochastic evolution,
since weak measurements under stochastic evolution should behave similarly
to time averaging studied in this work.

\section{Acknowledgments}

I'd like to thank Marko Vojinovi\'{c}, Nikola Paunkovi\'{c}, Igor
Salom, Aleksandra Go\v{c}anin, \v{C}aslav Brukner and Mihailo \DJ or\dj evi\'{c}
for useful discussions. Research supported by the Ministry of Science, Technological Development and Innovations (MNTRI) of the Republic of Serbia.

\end{document}